\documentclass[aps,prev,twocolumn,preprintnumbers,floatfix,nofootinbib]{revtex4-1}
\usepackage{graphicx}
\usepackage{amsmath, amsthm, amssymb}
\usepackage{amsfonts}
\usepackage{subfig}
\usepackage{xspace}
\usepackage{paralist}
\usepackage{multirow}
\usepackage{paralist}
\usepackage{graphicx}
\usepackage{bm}
\usepackage{times}
\usepackage{slashed}
\usepackage{color}
\usepackage{epsfig}\usepackage{dcolumn}
\usepackage{color}
\def\br{\begin{eqnarray}}
\def\er{\end{eqnarray}}
\def\be{\begin{equation}}
\def\ee{\end{equation}}
\newcommand{\MEG}{\mu \rightarrow e \gamma}

\begin{document}

\title{Type I + II Seesaw Model in light of the New Neutrino Oscillation Measurements}

\author{Maria Aguilar$^{1,2}$}
\author{Juan Carlos Helo$^{3,4}$}
\author{Toshihiko Ota$^{3,4}$}
\author{Farinaldo S. Queiroz$^{1,2,3,4}$}
\author{David Suarez$^2$}
\author{Amanda Rodr\'iguez$^{1,2,5,6}$}

\email{farinaldo.queiroz@ufrn.br}

\affiliation{$^1$ Departamento de F\'isica te\'orica e Experimental, Universidade Federal do Rio Grande do Norte, 59078-970, Natal, Rio Grande do Norte, Brazil}
\affiliation{$^2$International Institute of Physics, Universidade Federal do Rio Grande do Norte, Campus Universit\'ario, Lagoa Nova, Natal-RN 59078-970, Brazil} 
\affiliation{$^3$Millennium Institute for Subatomic Physics at High-Energy Frontier (SAPHIR), Fernandez Concha 700, Santiago, Chile.} 
\affiliation{$^4$ Departamento de F\'isica, Facultad de Ciencias, Universidad de La Serena, Avenida Cisternas 1200, La Serena, Chile.}
\affiliation{$^5$ Departamento de F\'isica, Universidad T\'ecnica Federico Santa Mar\'ia, Casilla 110-V, Valparaíso, Chile.}
\affiliation{$^6$ Instituto de F\'isica, Pontificia Universidad Cat\'olica de Valpara\'iso, Avenida Universidad 330, Curauma, Valpara\'iso, Chile}

\begin{abstract}
Global analysis of neutrino oscillation data slightly favors normal mass ordering. In this work, we investigate an extended scalar sector that naturally gives rise to a type I + II seesaw mechanism after spontaneous symmetry breaking and explore the interplay between collider physics and lepton flavor violation, adopting normal ordering. In particular, we focus on the rare muon decays $\mu \rightarrow e \gamma$ and $\mu \rightarrow 3e$ and the same-sign dilepton searches at LHC, a canonical signature of a doubly charged scalar. We conclude that neither the precise value of the sum of the neutrino masses, taken from DESI data that favors $\sum m_\nu=0.07$~eV, nor alternative cosmological fits which prefer a more relaxed limit $\sum m_\nu=0.1$~eV, significantly changes the theoretical prediction for these rare decays. However, we observe an interesting interplay between collider physics and lepton flavor violation.  In particular, we find that $\mu \rightarrow 3e$ is more constraining than $\mu \rightarrow e\gamma$. The $\mu \rightarrow 3e$ decay can yield a lower mass limit of $3$~TeV on the doubly charged scalar, surpassing current LHC constraints in a dominant type II seesaw or in a mixed type I+II scenario. Nevertheless, LFV does not strongly constrain the dominant type I seesaw scenario.
\end{abstract} 

\maketitle

\section{Introduction}

Although the Standard Model (SM) is a successful theory, it remains incomplete as it is not able to explain many phenomena. For example, astrophysical and cosmological data require the existence of a viable dark matter candidate~\cite{1980ApJ...238..471R,Planck:2018vyg}.
Neutrino oscillation experiments demand a viable mechanism to explain neutrino masses and mixing patterns~\cite{deSalas:2020pgw,Capozzi:2025wyn,Esteban:2024eli,ParticleDataGroup:2024cfk}.
Furthermore, the SM does not answer fundamental questions, e.g., 
the charge quantization~\cite{deSousaPires:1998jc,Dong:2005ebq}, the strong CP problem~\cite{Peccei:1977hh,Peccei:1977ur}, the number of fermion generations~\cite{Pisano:1996ht}, and the flavor structure for both the quark and lepton sectors~\cite{Feruglio:2015jfa,Novichkov:2021evw}.
Therefore, it is worthwhile to explore the possibilities of physics beyond the Standard Model that address some of these mysteries.

The existence of masses of neutrinos suggests that the lepton flavor violation (LFV) processes are allowed in contrast to the SM prediction. 
Those processes, of which no signal has been detected yet, are a powerful probe into physics beyond the SM, and allow constraining the parameter space of the models of this kind.

Majorana neutrino masses are provided in various models at tree level via the realization of a dimension five Weinberg operator~\cite{Weinberg:1979sa} that leads to Seesaw mechanisms of type I~\cite{Schechter:1980gr,Minkowski:1977sc,Yanagida:1980xy}, type II~\cite{Cheng:1980qt,Mohapatra:1979ia,Mohapatra:1999zr,Montero:2000rh,Ashanujjaman:2021txz}, and type III~\cite{Foot:1988aq,Montero:2001ts,Dorsner:2006fx,FileviezPerez:2008sr}. 
The type-I seesaw mechanism introduces heavy right-handed neutrinos to the SM field content, which are singlets under the SM gauge groups. 
In contrast,
a triplet scalar is introduced in the type-II seesaw mechanism, which acquires a vacuum expectation value (vev).
This mechanism yields a rich phenomenology due to the presence of new charged scalars. 
The type-III seesaw mechanism introduces a triplet of heavy fermions. 
The possibilities in which the type I and type II mechanisms are combined have been discussed to enrich the phenomenology~\cite{Akhmedov:2006de,Akhmedov:2008tb,Cogollo:2019mbd,Camargo:2018uzw}.

Instead of exploring the parameter space of the type I + II seesaw models in a simplified framework, in the present study, we address a full model, where the type I + type II mechanism emerges naturally. 
We are particularly interested in the model with the  $SU(3)_C\otimes{S}U(3)_L\otimes{U}(1)_N$ gauge symmetry~\cite{Singer:1980sw,Frampton:1992wt,Pisano:1992bxx,Foot:1992rh,Montero:1992jk,Foot:1994ym}, 3-3-1 for short.
The model can nicely explain the reason why the number of fermion generations is three; The gauge anomalies are absent only if there is a multiple of three families in their fermion spectrum, and there is no room for more than three fermion generations with the consideration of the asymptotic freedom \cite{Dias:2004dc,Dias:2004wk}. 
With the introduction of three $SU(3)_{L}$-triplet scalars, we can generate masses for SM fields except for neutrinos.

The right-handed neutrinos are introduced by extending the SM lepton doublet into a triplet of $SU(3)_{L}$, which leads to the type-I seesaw mechanism \cite{Long:1995ctv,Long:1996rfd}. 
An alternative way to provide neutrino masses is the introduction of a $SU(3)_{L}$ sextet scalar, which resembles the introduction of a scalar triplet to generate the type-II seesaw mechanism \cite{Foot:1994ym,Ky:2005yq,Dong:2008sw,Ferreira:2019qpf}.

In this study, we show that the introduction of a scalar sextet interestingly gives rise to a combination of the type I and type II seesaw mechanisms. 
After spontaneous symmetry breaking (SSB), the scalar sextet breaks down to a scalar triplet and a scalar doublet, which yield the type I + II seesaw mechanism together with the right-handed neutrinos in the $SU(3)_L$ triplet fermions.
We evaluate the size of relevant LFV processes and derive the bounds on the model, taking into account the most up-to-date results of the global fit of neutrino oscillation parameters \cite{deSalas:2020pgw,Capozzi:2025wyn,Esteban:2024eli,ParticleDataGroup:2024cfk}, which weakly favor normal mass ordering.
%
%
We also take into account the cosmological constraint on neutrino masses, which was recently updated with the measurement of the baryon acoustic oscillation by the DESI Collaboration \cite{DESI:2024mwx}.
It imposed the sum of neutrino masses to be $\sum m_\nu <  0.072$~eV \cite{DESI:2024mwx}. In this work, we conservatively adopt $\sum m_\nu=0.07$ eV to be consistent with Planck and
DESI measurements \cite{Planck:2018vyg,DESI:2024mwx} and also investigate the impact of this value on our conclusions.

This work extends previous studies by providing a fresh perspective to the 3-3-1 model, which brings the type I + type II seesaw mechanisms with an $SU(3)_{L}$ sextet scalar,
taking into account the bounds of the LFV processes in light of new data, which differs from the existing literature \cite{Foot:1994ym,Dong:2008sw,Pinheiro:2022obu,DeConto:2015eia}.

The doubly and singly charged scalars that arise from the scalar sextet and the relative contribution of the scalar vevs of the neutral fields within the sextet drive the LFV phenomenology. 
Thus, we report our results in terms of their masses. 
In parallel to that, the doubly charged Higgs bosons that decay promptly into same-sign charged leptons have been studied in the context of the hadron collider searches \cite{Chen:2008qb,Crivellin:2018ahj,Chaudhuri:2013xoa,ATLAS:2018ceg,deMelo:2019asm,LHeC:2020van,Enomoto:2021fql,Ducu:2024xxf,Bolton:2024thn,Zhou:2025ljo,FCC:2025lpp,FCC:2025uan,FCC:2025jtd,Chiang:2025lab}. 
Therefore, we compare our findings with the most recent LHC results using proton–proton collision
at a center-of-mass energy of 13 TeV and $36.1 fb^{-1}$ of integrated luminosity \cite{ATLAS:2017xqs}. 
Our goal is to investigate the interplay between LFV and collider physics in the framework of the 3-3-1 model. 

This paper is organized as follows. Section~\ref {Model_overview} provides an overview of the model with a $SU(3)_C\otimes{S}U(3)_L\otimes{U}(1)_N$ symmetry with right-handed neutrinos and type I + type II Seesaw. Sec.~\ref{Lepton_Flavor_Violation} analyzes constraints from LFV processes in perspective with collider bounds. Finally, we conclude in Sec.~\ref{Conclusions}.

\section{Model overview}
\label{Model_overview}


In the $SU(3)_C\otimes{S}U(3)_L\otimes{U}(1)_N$ model, 3-3-1 for short, the left-handed chiral fermions in this model are triplets in a fundamental representation of $SU(3)_L$, while the right-handed fermions remain as singlets. For example, left-handed leptons are $SU(3)_L$ triplets
\begin{equation}
\psi_{aL}=\begin{pmatrix}
\nu_a\\
e_a\\
N^c_a
\end{pmatrix}_L\sim(\mathbf{1},\mathbf{3},-1/3)\,,
\end{equation}
where $a=1,2,3$. The third component of the triplet corresponds to a right-handed neutrino. However, right-handed leptons are $SU(3)_L$ singlets
\begin{equation}
e_{aR}\sim(\mathbf{1},\mathbf{1},-1)\,.    
\end{equation}
Notice that in this model, each lepton family is treated democratically. 

In the quark sector, however, the first two quark families are treated differently from the third family to cancel out gauge anomalies. Left-handed quarks of the first two families are $SU(3)_L$ anti-triplets
\begin{equation}
Q_{\alpha{L}}=\begin{pmatrix}
d_{\alpha}\\
-u_{\alpha}\\
D_{\alpha}
\end{pmatrix}_L\sim(\mathbf{3},\mathbf{3}^*,0)\,,    
\end{equation}
where $\alpha=1,2$, while the third generation of left-handed quarks is a $SU(3)_L$ triplet
\begin{equation}
Q_{3{L}}=\begin{pmatrix}
u_3\\
d_3\\
U
\end{pmatrix}_L\sim(\mathbf{3},\mathbf{3},1/3)\,,    
\end{equation}
The right-handed quarks transform as
\begin{align}
u_{\alpha{R}}\sim & (\mathbf{3},\mathbf{1},2/3)\,,\\
d_{\alpha{R}}\sim & (\mathbf{3},\mathbf{1},-1/3)\,,\\
D_{\alpha{R}}\sim & (\mathbf{3},\mathbf{1},-1/3)\,,\\
u_{3R}\sim & (\mathbf{3},\mathbf{1},2/3)\,,\\
d_{3R}\sim & (\mathbf{3},\mathbf{1},-1/3)\,,\\
U_R\sim & (\mathbf{3},\mathbf{1},2/3)\,.
\end{align}

The mass spectrum for the charged fermions in the model is generated via the following three scalar triplets
\begin{align}
\chi= & \begin{pmatrix}
\chi^0_1\\
\chi^-_2\\
\chi^0_3
\end{pmatrix}\sim(\mathbf{1},\mathbf{3},-1/3)\,,\\
\eta= & \begin{pmatrix}
\eta^0_1\\
\eta^-_2\\
\eta^0_3
\end{pmatrix}\sim(\mathbf{1},\mathbf{3},-1/3)\,,\\
\rho= & \begin{pmatrix}
\rho^+_1\\
\rho^0_2\\
\rho^+_3
\end{pmatrix}\sim(\mathbf{1},\mathbf{3},2/3)\,.
\end{align}
When a neutral component of a scalar triplet $\chi$ acquires a vev, $SU(3)_L\otimes{U}(1)_N$ is spontaneously broken into $SU(2)_L\otimes{U}(1)_Y$ to guarantee the proper low-energy phenomenology. After that, the vev of a neutral component of $\eta$ and $\rho$ spontaneously breaks $SU(2)_L\otimes{U}(1)_Y$ into $U(1)_Q$. In other words, spontaneous symmetry breaking proceeds as follows
\begin{equation}
SU(3)_L\otimes{U}(1)_N\xrightarrow{\langle\chi\rangle}SU(2)_L\otimes{U}(1)_Y\xrightarrow{\langle\rho\rangle,\langle\eta\rangle}U(1)_Q \,.   
\end{equation}
We consider here a scenario where the scalar fields have a broken phase
\begin{align}
\chi= & \frac{1}{\sqrt{2}}\begin{pmatrix}
0\\
0\\
v_{\chi}
\end{pmatrix}\,,\\
\eta= & \frac{1}{\sqrt{2}}\begin{pmatrix}
v_{\eta}\\
0\\
0
\end{pmatrix}\,,\\
\rho= & \frac{1}{\sqrt{2}}\begin{pmatrix}
0\\
v_{\rho}\\
0
\end{pmatrix}\,.
\end{align}
From now on, we suppose $v_{\eta}=v_{\rho}$ and $v^2_{\eta}+v^2_{\rho}\approx{246}^2\mathrm{GeV}^2$ to simplify our calculations and guarantee the right value for the $W$ and $Z$ boson masses.

Additionally, we impose an extra $Z_2$ symmetry that avoids mixing terms between SM and exotic quarks and dangerous flavor-changing neutral currents.  That said, fields transform under $Z_2$ as
\begin{align}
u_{aR}\to & -{u}_{aR}\,,\\
d_{aR}\to & -{d}_{aR}\,,\\
\eta\to & -\eta\,,\\
\rho\to & -\rho\,,\\
e_{aR}\to & -{e}_{aR}\,,
\end{align}
while the other fields transform trivially under this $Z_2$ symmetry. Consequently, the allowed term for the Yukawa Lagrangian that generates masses for the charged leptons is
\begin{equation}
\mathcal{L}^l_Y=h^l_{ab}\overline{\psi}_{aL}\rho{e}_{Rb}+\mathrm{h.c}\,,    
\end{equation}
where the mass matrix for charged leptons reads
\begin{equation}
M^l=h^l\frac{v_{\rho}}{\sqrt{2}}\,.
\end{equation}
Similarly, the Yukawa Lagrangian for quarks reads,
\begin{align}
\mathcal{L}^q_Y= & h^u_{\alpha{a}}\overline{Q}_{\alpha{L}}\rho^*u_{aR}+h^d_{\alpha{a}}\overline{Q}_{\alpha{L}}\eta^*d_{aR}+h^U\overline{Q}_{3L}\chi{U}_R\nonumber\\
& +h^d_{3a}\overline{Q}_{3L}\rho{d}_{aR}+h^u_{3a}\overline{Q}_{3L}\eta{u}_{aR}+h^D_{\alpha\beta}\overline{Q}_{\alpha{L}}\chi^*d_{\beta{R}}+\mathrm{h.c}\,,
\end{align}
and yields mass matrices
\begin{align}
M^u= & \frac{1}{\sqrt{2}}\begin{pmatrix}
-v_{\rho}h^u_{11} & -v_{\rho}h^u_{12} & -v_{\rho}h^u_{13} \\
-v_{\rho}h^u_{21} & -v_{\rho}h^u_{22} & -v_{\rho}h^u_{23} \\
v_{\eta}h^u_{31}  &  v_{\eta}h^u_{32} &  v_{\eta}h^u_{33}
\end{pmatrix}\,,\\
M^d= & \frac{1}{\sqrt{2}}\begin{pmatrix}
v_{\eta}h^d_{11} & v_{\eta}h^d_{12} & v_{\eta}h^u_{13} \\
v_{\eta}h^d_{21} & v_{\eta}h^d_{22} & v_{\eta}h^u_{23} \\
v_{\rho}h^u_{31} & v_{\rho}h^u_{32} & v_{\rho}h^u_{33}
\end{pmatrix}\,,\\
M^U= & \frac{1}{\sqrt{2}}v_{\chi}h^U\,,\\
M^D= & \frac{1}{\sqrt{2}}\begin{pmatrix}
v_{\chi}h^D_{11} & v_{\chi}h^D_{12} \\
v_{\chi}h^D_{21} & v_{\chi}h^D_{22}
\end{pmatrix}\,.
\end{align}
New exotic quarks $D_{\alpha}$ and $U$ can be produced at the LHC via Drell-Yan processes. We assume them to be very heavy to avoid collider constraints~\cite{ATLAS:2024kgp,Benbrik:2024fku}. In this model, we can achieve this by choosing a large scale for SSB of ${S}U(3)_L\otimes{U}(1)_N$ symmetry.

We will not revisit this here, but it is well-known that such scalar triplets generate masses to all fermions except for the neutrinos. Interested in generating neutrino masses, we add a scalar sextet with ~\cite{Foot:1994ym,Ky:2005yq,Dong:2008sw,Ferreira:2019qpf}
\begin{equation}
S=\begin{pmatrix}
S^0_{11} & S^-_{12}    & S^0_{13} \\
S^-_{12} & S^{--}_{22} & S^-_{23} \\
S^0_{13} & S^-_{23}    & S^0_{33}
\end{pmatrix}\sim(\mathbf{1},\mathbf{6},-2/3)\,.
\end{equation}
In the broken phase, it reads,
\begin{equation}
\langle{S}\rangle=\frac{1}{\sqrt{2}}\begin{pmatrix}
v_{s11} & 0 & v_{s13} \\
0       & 0 & 0 \\
v_{s13} & 0 & \Lambda
\end{pmatrix}\,.
\end{equation}
After $S_{33}^0$ gets a vev, $\Lambda$, the scalar sextet breaks into a $SU(2)_L$ triplet, a $SU(2)_L$ doublet and a $SU(2)_L$ singlet
\begin{equation}
S\to{S}_1(\mathbf{1},\mathbf{3},-2)+{S}_2(\mathbf{1},\mathbf{2},1)+S_3(\mathbf{1},\mathbf{1},0)\,,   
\end{equation}
where 
\begin{align}
S_1= & \begin{pmatrix}
S^0_{11} & S^-_{12} \\
S^-_{12} & S^{--}_{22}
\end{pmatrix}\,,\\
S_2= & \begin{pmatrix}
S^{0}_{13}\\
S^-_{23}
\end{pmatrix}\,,\\
S_3= & S^0_{33}\,.
\end{align}
gives a Majorana mass term for active neutrinos via the Yukawa Lagrangian
\begin{equation}
\mathcal{L}=f^{\nu}_{ab}(\overline{\psi}_{aL})_{m}(\psi^c_{bL})_{n} S_{mn}+\mathrm{h.c}\,.  
\end{equation}
After SSB, the Lagrangian can be rewritten as,
\begin{equation}
\mathcal{L}=\frac{1}{2}\begin{pmatrix}
\overline{\nu}_{aL} & \overline{N}^c_{aR}    
\end{pmatrix}\begin{pmatrix}
(M_L)_{ab} & (M_D)_{ab} \\
(M^T_D)_{ab} & (M_R)_{ab} 
\end{pmatrix}\begin{pmatrix}
\nu^c_{bL} \\
N_{bR}
\end{pmatrix}\,,
\end{equation}
where
\begin{align}
(M_L)_{ab} = & \sqrt{2}v_{s11}(f^{\nu})_{ab}\,,\\
(M_D)_{ab} = & \sqrt{2}v_{s13}(f^{\nu})_{ab}\,,\\
(M_R)_{ab} = & \sqrt{2}\Lambda(f^{\nu})_{ab}\,.
\end{align}
Then, the mass matrix for active neutrinos is found to be
\begin{equation}
(m^{\nu})_{ab}=\sqrt{2}\left(v_{s11}-\frac{v^2_{s13}}{\Lambda}\right)(f^{\nu})_{ab}\,,    
\end{equation}
and the mass matrix for right-handed neutrinos
\begin{equation}
(M^{N})_{ab}=\sqrt{2}\Lambda(f^{\nu})_{ab}\,.    
\end{equation}
We highlight that the scalar sextet gives rise to a scalar triplet, a doublet, and a singlet. The scalar doublet ($S_{13}^0, S_{23}^-$) generates a Dirac mass term for neutrinos proportional to $v_{s13}$ and the scalar triplet produces a Majorana mass term governed by $v_{s11}$. We emphasize that this model differs from the previous proposal presented in \cite{Dong:2008sw}. The right-handed neutrinos have masses set by the energy scale $\Lambda$. Hence, before SSB, neutrinos are massless, but after the scalar text develops a vev, a type I + II seesaw emerges naturally. To simplify the phenomenological analysis, we consider the following assumption
\begin{equation}
\frac{v^2_{s13}}{\Lambda}=k{v_{s11}}\,,    
\end{equation}
therefore the neutrino mass matrix reads
\begin{equation}
(m^{\nu})_{ab}=\sqrt{2}(1-k)v_{s11}(f^{\nu*})_{ab}\,,    
\end{equation}
I define
\begin{equation}
\kappa\equiv{1}-k\,,    
\end{equation}
then the neutrino mass matrix simplifies to
\begin{equation}
\label{eqnumasses}
(m^{\nu})_{ab}=\sqrt{2}\kappa{v}_{s11}(f^{\nu*})_{ab}\,.    
\end{equation}

Here, we can distinguish three scenarios. First, when $|\kappa|\gg{1}$, we have a dominant type I seesaw. In the case of $|\kappa|\sim{1}$, we have a dominant type II seesaw. Finally, in the case of $|\kappa|\sim{0}$, we have a full type I+II seesaw, with both processes making significant phenomenological contributions. 

The gauge sector will not drive our LFV phenomenology, but we will cover it because it yields an important bound on $v_{s11}$. The gauge bosons of the $SU(3)_L\otimes{U}(1)_N$ symmetry acquire masses via SSB of the scalar triplets. To illustrate this, we consider the covariant derivative of $SU(3)_L\otimes{U}(1)_N$ symmetry 
\begin{equation}
D_{\mu}X=\left(\partial_{\mu}-ig_LW^m_{\mu}\frac{\lambda^m}{2}-ig_NN_{X}W^N_{\mu}\right)X\,,
\end{equation}
where $g_L$ is the gauge coupling from $SU(3)_L$ symmetry, $W^{m}$ are the eight gauge bosons in the adjoint representation of $SU(3)_L$, $\lambda^m$ are the Gell-Mann matrices, $g_N$ is the gauge coupling from $U(1)_N$ symmetry, and $W^N$ is the gauge boson of $U(1)_N$ symmetry. When the covariant derivative acts on the scalar fields that acquire vev, $\chi$, $\rho$, $\eta$, and $S$, it renders terms that contribute to the masses of gauge bosons. Explicitly, the term proportional to $\lambda^m$ can be written as
\begin{equation}
\frac{g_L}{2}W^m_{\mu}\lambda^m=\frac{g_L}{2}\begin{pmatrix}
W^3_{\mu}+\frac{1}{\sqrt{3}}W^8_{\mu} & \sqrt{2}W^{-}_{\mu}                    & \sqrt{2}U^0_{\mu}           \\    
\sqrt{2}W^{+}_{\mu}                   & -W^3_{\mu}+\frac{1}{\sqrt{3}}W^8_{\mu} & \sqrt{2}W^{'+}_{\mu}        \\
\sqrt{2}{U^{0}_{\mu}}^{*}             & \sqrt{2}W^{'-}_{\mu}                   & \frac{2}{\sqrt{3}}W^8_{\mu}     
\end{pmatrix}\,,    
\end{equation}
where:
\begin{align}
W^{\pm}_{\mu}= & \frac{1}{\sqrt{2}}(W^1_{\mu}\mp{i}W^2_{\mu})\,,\\
U^0_{\mu}= & \frac{1}{\sqrt{2}}(W^4_{\mu}-iW^5_{\mu})\,,\\
W^{'\pm}_{\mu}= & \frac{1}{\sqrt{2}}(W^6_{\mu}\pm{i}W^7_{\mu})\,.
\end{align}
The bosons $W^3_{\mu}$, $W^8_{\mu}$, and $W^N_{\mu}$ mix to form three mass eigenstates: photon, SM $Z$ boson, and a new $Z'$ boson. Moreover, we find an additional neutral boson $U^0_{\mu}$~\cite{RamirezBarreto:2008zz} and a charged gauge boson $W^{'+}_{\mu}$. Collider searches for those gauge bosons have been done and result in bounds at the $4-5$~TeV energy scale~\cite{Salazar:2015gxa,Lindner:2016lpp,Arcadi:2017xbo,DeJesus:2020yqx,Alves:2022hcp,deJesus:2023lvn,deJesus:2023som}.

Regarding the mass spectrum of gauge bosons, the $W$ and $Z$ bosons receive contributions from vevs of the scalar sextet, specifically, from $v_{s11}$ and $v_{s13}$. Therefore, the constraint of the electroweak precision test on the parameter $\rho$ imposes the constraint~\cite{Camargo:2018uzw}
\begin{equation}
v_{s11}<2\,\mathrm{GeV}\,.    
\end{equation}
This limit is important for our reasoning because we parametrize our analysis in terms of the vev $v_{s11}$ and $\kappa$. We will set $v_{s11} \sim 1$~eV as the benchmark.

The masses of the new gauge bosons $Z'$ and $W'$ read
\begin{align}
M^2_{Z'}\approx&\frac{g^2_L}{4h_W\cos^2\theta_W}\left[v^2_{\rho}+2v^2_{s13}+(v^2_{\eta}+4v^2_{s11})\cos^2(2\theta_W)\right.\nonumber\\
&\left.+4(v^2_{\chi}+4\Lambda^2)\cos^4 \theta_W\right]\\
M^2_{W'}\approx&\frac{g^2_L}{4}\left(v^2_{\rho}+v^2_{\chi}+4v^2_{s13}+2\Lambda^2\right)\,.
\end{align}
where
\begin{equation}
\cos^2 \theta_W=\frac{3+t^2}{3+4t^2}\,,    
\end{equation}
and
\begin{equation}
h_W=3-4\sin^2\theta_W\,,    
\end{equation}
with $t=g_N/g_L$. Now that we have reviewed the key ingredients of the proposed type I+II seesaw model, we will address the relevant collider bounds.

\section{Collider Bounds}
\label{Collider_Bounds}

The search for a 3-3-1 symmetry has been done in terms of $Z^\prime$ and $W^\prime$ signals. The non-abelian nature of the 3-3-1 extension gives rise to both a $Z^\prime$ and $W^\prime$ boson. Their masses are directly connected. Studies have derived lower mass bounds on the $Z^\prime$ boson. The most updated limit imposes $M_{Z^\prime} > 4-5$~TeV.  However, one can assume a 3-3-1 symmetry breaking scale, $v_\chi$, to be very high and avoid these limits. Moreover, our LFV phenomenology is not bounded by this constraint because the mass of the doubly charged scalar is governed by the bilinear term $\mu_S^2 Tr(S^\dagger S)$, not $v_\chi$. One can tune $\mu_S^2$ to be sufficiently large to increase the mass of the doubly charged scalar \cite{Pinheiro:2022obu}. Hence, the important collider bound stems from the same-sign dilepton search at the LHC, which is a clear signature of a doubly charged scalar. The limit relies on the Drell-Yan production of pairs of doubly charged scalars. Thus, it depends only on the decaying final states. As electrons, muons, and taus are subject to different detector efficiencies, the limit changes a bit according to the final leptons. In our work, we will adopt the limit $M_{S^{\pm\pm}}> 856$~ GeV \cite{ATLAS:2017xqs} found for a doubly charged scalar decaying mainly into muon pairs (53\%), and only (1\%) into electrons, which is our case. We exhibit it in the FIGs.\ref{fig:result1}-\ref{fig:result2}.

Having presented the relevant interactions and the collider bound for our reasoning, we address in the next section the LFV signatures.

\section{Lepton Flavor Violation}
\label{Lepton_Flavor_Violation}

The LFV processes are one of the most powerful probes for physics beyond the SM, because they are strictly prohibited in the SM and also strongly suppressed in the SM extended with neutrino masses.
In this study, we focus on $\mu \rightarrow e \gamma$ and $\mu \rightarrow 3e$ processes.
It is known that the $\mu$-$e$ conversion in nuclei and the LFV processes associated with tau flavour only provide subdominant bounds to the parameters of the model discussed in this paper~\cite{Lychkovskiy:2010ue,Dinh:2012bp,Abada:2014kba,He:2014efa}. 
We evaluate the branching ratios of the $\mu \rightarrow e \gamma$ and $\mu \rightarrow 3e$ processes and discuss how the sensitivities to the model parameters change under the choice of the neutrino mass parameters and the vacuum expectation value $v_{s11}$. 
We also compared the results with the current limits set by the LHC.
\begin{table}[t]
\begin{equation*}
\begin{array}{|c|c|c|c|}
\hline
 & \text{Benchmark \#1} & \text{Benchmark \#2}\\
\hline
\Delta{m}^2_{21}/(10^{-5}\rm{eV}^2) & 7.41 & 7.37\\
\sin^2\theta_{12}/10^{-1}           & 3.03 & 3.03\\
\Delta{m}^2_{32}/(10^{-3}\rm{eV}^2) & 2.437 & 2.495\\
\sin^2\theta_{13}/10^{-2}           & 2.203 & 2.23 \\
\sin^2\theta_{23}/10^{-1}           & 5.72 & 4.73 \\
\delta/\pi                          & 1.094 & 1.2\\
\hline
\end{array}   
\end{equation*}
\caption{Two sets of our benchmark choices of neutrino mass and mixing parameters.
\#1 is the best-fit parameters taken from \cite{Workman:2022ynf} and \# 2 from \cite{Capozzi:2025wyn}, which correspond to the two choices in the $\theta_{23}$ octant.}
\label{tab:numixing}
\end{table} 
The current best limits for these processes are $\text{BR}(\mu \rightarrow e \gamma)<1.5\times 10^{-13}$ at $90\%\,\, \text{C}.\text{L}$ obtained by the MEGII Collaboration~\cite{MEGII:2025gzr} and $\text{BR}(\mu \rightarrow 3e)< 10^{-12}$ set by SINDRUM~\cite{SINDRUM:1987nra}. 
However, those bounds are expected to be improved up to $\text{BR}(\mu \rightarrow e \gamma)<6\times 10^{-14}$ and $\text{BR}(\mu \rightarrow 3e)< 10^{-16}$~\cite{COMET:2025sdw,Palo:2025oyq}. 

In the scenario with a type I and type II seesaw contribution to neutrino masses,
\begin{equation}
\text{BR}(\mu \rightarrow e\gamma )\simeq \frac{\alpha _{EM}\left\vert
 (f^{\nu \dagger}f^{\nu})_{e\mu }\right\vert ^{2}}{192\pi G_{F}^{2}\text{
}}\left( \frac{1}{M_{S^{\pm\pm}}^{2}}+\frac{8}{M_{S ^{\pm}}^{2}}\right) ^{2}\,,
\label{eqBmue}
\end{equation}
where $\alpha_{EM}$ is the fine-structure constant, $G_F$ the Fermi constant, $M_{S^{\pm\pm}}$ and $M_{S^{\pm}}$ the masses of the doubly charged and singly charged scalar, respectively, and $f^{\nu}_{ab}$ can be given with the neutrino mass and mixing parameters as
\begin{equation}
\label{eq:Yukawa}
f^{\nu}_{ab}=\frac{1}{\sqrt{2}\kappa{v}_{s11}} U^{\ast}_{ai} \text{diag}(m_{\nu1},m_{\nu2},m_{\nu3})_{i} U^{\dagger}_{ib}\,,
\end{equation}
where $U$ is the Pontecorvo-Maki-Nakagawa-Sakata (PMNS) lepton mixing matrix~\cite{PhysRevD.110.030001, Dinh:2012bp}, which is parameterized as
\begin{align}
U=&\left(\begin{array}{ccc}
1 & 0   & 0 \\
0  & \cos\theta_{23} & \sin\theta_{23} \\
0 & -\sin\theta_{23}  & \cos\theta_{23}
\end{array}  \right) \left(\begin{array}{ccc}
\cos\theta_{13} & 0   & \sin\theta_{13} e^{-i\delta} \\
0  & 1 & 0 \\
-\sin\theta_{13} e^{i\delta} & 0 & \cos\theta_{13}
\end{array}  \right)\nonumber\\
&\times\left(\begin{array}{ccc}
\cos\theta_{12} & \sin\theta_{12}   & 0 \\
-\sin\theta_{12}  & \cos\theta_{12} & 0 \\
0 & 0  & 1
\end{array}  \right)\,.
\end{align}
In short, the LFV processes are directly related to neutrino observables through the Yukawa couplings $f^{\nu}_{ab}$. 
Both the singly and the doubly charged scalars contribute to the $\mu \rightarrow e \gamma$ process, and the ratio of those two contributions is determined with their masses as shown in Eq.\eqref{eqBmue}. 
Differing from $\mu \rightarrow e \gamma$, only the doubly charged scalar field mediates the $\mu \rightarrow 3e$ process,
and its branching ratio is found to be
\begin{equation}
\text{BR}(\mu \rightarrow 3e) =\frac{\left\vert
 (f^{\nu \dagger}f^{\nu})_{e\mu }\right\vert ^{2} }{G_F^2 M_{S^{\pm\pm}}^4}\, .
\label{eqBR3e}
\end{equation}

\begin{table}[t]
\centering
\begin{tabular}{|c|c|c|c|c|}
\hline
cf. Tab. I
& $\kappa$ & BR ($\MEG$) & ${\rm BR} (\mu \rightarrow 3e)$
\\
\hline 
\#1 & $0.1$    & $1173.47/M_{S^{\pm\pm}}^4$& $ 1.23\, \cdot\,10^6/M_{S^{\pm\pm}}^4$
    \\
    & $1$    & $0.1/M_{S^{\pm\pm}}^4$& $ 123.5/M_{S^{\pm\pm}}^4$ \\
    & $10$    & $1\cdot 10^{-5}/M_{S^{\pm\pm}}^{4}$& $ 0.012/M_{S^{\pm\pm}}^4$
    \\
\hline
\#2 
& $0.1$ & $1125.7/M_{S^{\pm\pm}}^{4}$ & $1.14\, \cdot\, 10^6/M_{S^{\pm\pm}}^{4}$
\\
& $1$ & $0.11/M_{S^{\pm\pm}}^{4}$ & $114.87/M_{S^{\pm\pm}}^{4}$
\\
& $10$ & $1.12\, \cdot\, 10^{-5}/M_{S^{\pm\pm}}^{4}$ & $0.011/M_{S^{\pm\pm}}^{4}$
\\

\hline
\end{tabular}
\caption{Branching ratios calculated for several values of $\kappa$ and the two benchmark sets of neutrino oscillation parameters in Tab.~\ref{tab:numixing}.} 
\label{tab:dataset}
\end{table}

In this study, we assume that the doubly charged scalar has a mass with the same size as that of the singly charged one, allowing us to directly connect the LFV observables with the collider searches for doubly charged scalars.
As shown in Eq.\eqref{eqBmue} and Eq.\eqref{eqBR3e}, the $\mu \rightarrow e \gamma$ and $\mu \rightarrow 3e$ decays depend on the particular combinations of the Yukawa couplings that are directly related to the observed neutrino masses, cf. Eq.\eqref{eqnumasses}. 
Therefore, the size of the LFV decays is strongly tied to the neutrino mass spectrum and parameters. 
In this study, we take the normal hierarchy for the neutrino mass ordering as a reference, which is slightly preferred in the global fits of the results of the oscillation experiments \cite{deSalas:2020pgw,Esteban:2024eli,Capozzi:2025wyn}.
%
%
We use two sets of neutrino oscillation parameters to compute the branching ratio for both the LFV processes.
The first set is taken from the 2024 Particle Data Group (PDG)~\cite{ParticleDataGroup:2024cfk}, and the second one is from the review~\cite{Capozzi:2025wyn}, which are presented in Tab.~\ref{tab:numixing}.\footnote{In \cite{Capozzi:2025wyn}, the mass-square difference corresponding to the atmospheric neutrino oscillation is defined as $\Delta m^{2} = \Delta m_{32}^{2} + \Delta m_{21}^{2}/2$. However, the difference does not make any significant change in our results of the calculations of the LFV processes.} 
These two benchmark parameter sets correspond to the two choices in the $\theta_{23}$ octant, which have not been determined by the global fit analyses.
Observations of CMB by Planck~\cite{Planck:2018vyg}, BAO-DESI~\cite{DESI:2024mwx} and the CMB lensing by ACT~\cite{ACT:2025tim} point to $\sum m_{\nu} < 0.11$ eV with the prior of normal hierarchy. Therefore, we choose $m_{\nu1}=8\, \text{meV}$ to be consistent with them.
%
%

\begin{figure}[ht] 
    \centering
\includegraphics[width=0.45\textwidth]{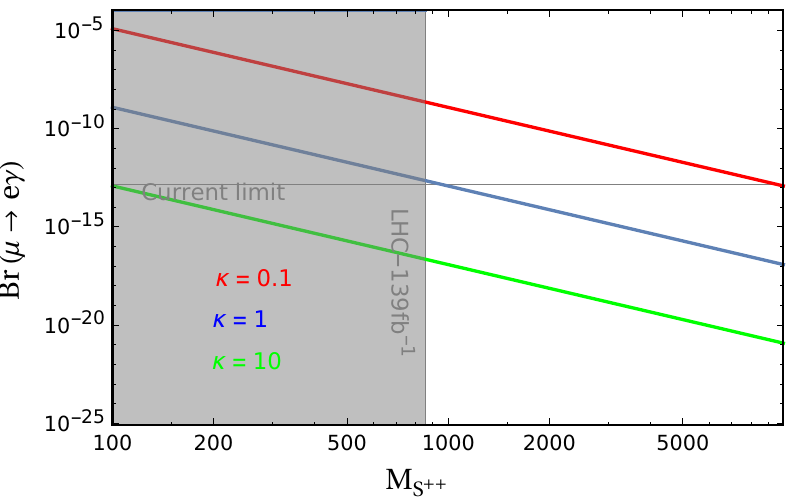}
\includegraphics[width=0.45\textwidth]{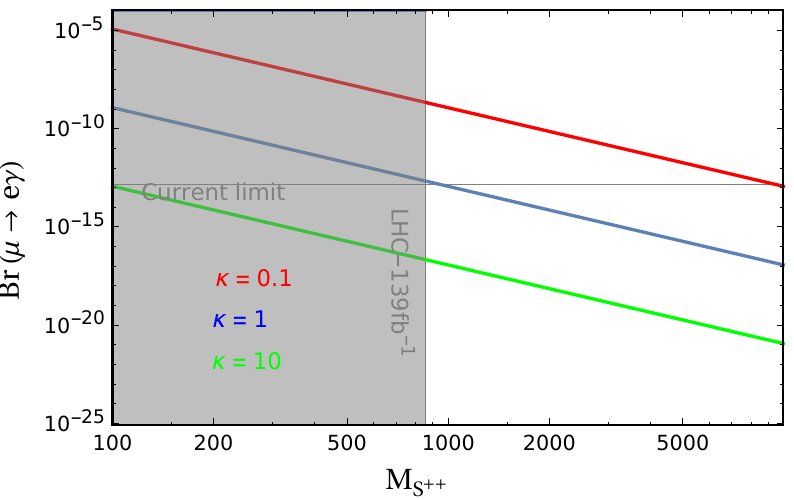} 
    \caption{Branching ratio for $\mu \rightarrow e \gamma$ decay with several values of $\kappa$ parameter: $\kappa=0.1$ (red), $\kappa=1$ (blue) and $\kappa=10$ (green). The upper (lower) plot uses the oscillation-parameter values from benchmark \#1 (\#2), as shown in Tab.~\ref{tab:numixing}.}
    \label{fig:result1}
\end{figure}

\begin{figure}[t] 
    \centering
\includegraphics[width=0.45\textwidth]{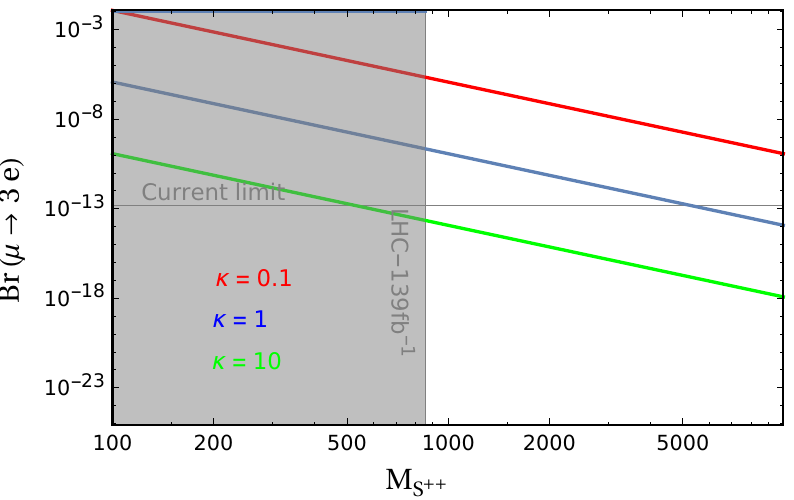}
\includegraphics[width=0.45\textwidth]{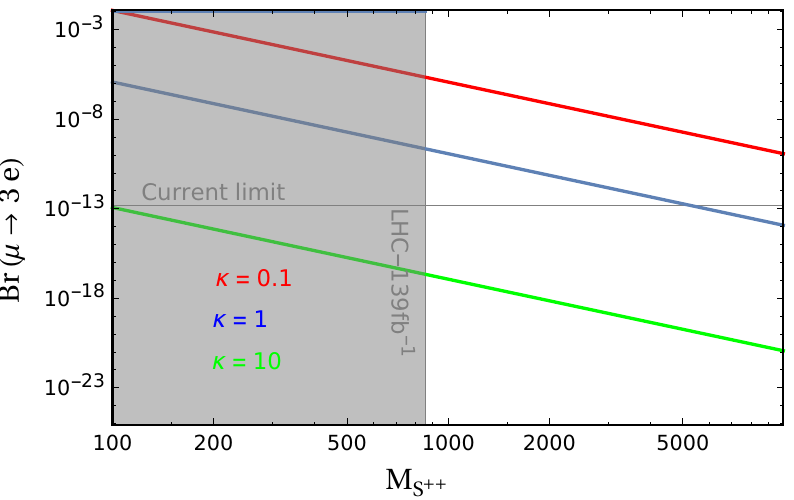} 
    \caption{Branching ratio for $\mu \rightarrow 3e$ decay with several values of $\kappa$ parameter: $\kappa=0.1$ (red), $\kappa=1$ (blue) and $\kappa=10$ (green). The upper (lower) plot uses the oscillation parameter values from benchmark \#1 (\#2), as shown in Tab.~\ref{tab:numixing}.}
    \label{fig:result2}
\end{figure}


In Tab.~\ref{tab:dataset}, we show the branching ratios of the LFV processes as functions of the mass of the doubly charged scalar field for some values of the $\kappa$ parameter, using our benchmark parameter choices.
Here the vacuum expectation value $v_{s11}$ of the sextet scalar field is fixed at $v_{s11}=1\, \text{eV}$. The branching ratios depend on the explicit choice of the $\kappa$ parameter as a direct result of Eq. \eqref{eq:Yukawa}. The branching ratios for both the LFV processes are plotted as functions of $M_{S^{\pm\pm}}$ for some values of the $\kappa$ parameter in FIGs.~\ref{fig:result1}-\ref{fig:result2}.
The shaded regions are excluded by the same-sign dilepton searches at the LHC as discussed in the previous section.
%


%
The bound from SINDRUM~\cite{SINDRUM:1987nra} to $\mu \rightarrow 3e$ in FIG.~\ref{fig:result2} is delimited by a horizontal line. 
In all plots, we show the theoretical predictions for the branching for several values of the $\kappa$ parameter: $\kappa=0.1$ (red), $\kappa=1$ (blue), and $\kappa=10$ (green). The most restricted scenario is where both type I and type II contribute significantly to LFV (red curve), while the least constrained scenario is the one with a dominant type I seesaw. For both the dominant type II seesaw and the mixed scenario, we can conclude that $\mu \rightarrow e\gamma$ provides a stronger constraint than the LHC, as seen in FIG.~\ref{fig:result1}. 
This conclusion holds regardless of the choice of the neutrino oscillation parameters given in Tab.~\ref{tab:numixing}. This finding is anticipated in Tab.~\ref{tab:dataset}, which shows a small difference in the predicted branching ratios. For a dominant type I seesaw, the only constraint comes from the collider limit on the masses of charged scalars, since LFV is far beyond experimental sensitivity. This clearly proves the importance of continuing to search for LFV signals in colliders.  

Concerning the $\mu \rightarrow 3e$ decay, most of the arguments for $\mu \rightarrow e\gamma$ stay the same. However, this decay mode places a much stronger constraint than $\mu \rightarrow e\gamma$ reaching $M_{S^{\pm\pm}} > 3$~TeV for $\kappa=0$ and $\kappa=1$ (for a dominated type II scenario and a mix type I+II scenario). We highlight that two choices of the neutrino oscillation parameters do not make a visible change in the branching ratios. In summary, it is clear that $\mu \rightarrow 3e$ is and will be the best lepton flavor violating probe for this model. The rich interplay between LFV and collider physics will continue in light of the upcoming results from $\mu \rightarrow 3e$ searches and the future colliders~\cite{LHeC:2020van,MuonCollider:2022xlm,Accettura:2023ked,MuCoL:2024oxj,InternationalMuonCollider:2024jyv,FCC:2025lpp,FCC:2025uan}. 
%



\section{Conclusions}
\label{Conclusions}

We investigate the LFV predictions in a $SU(3)_C\otimes{S}U(3)_L\otimes{U}(1)_N$ model with right-handed neutrinos featuring a type I + type II Seesaw, having in mind updated neutrino oscillation experiments that slightly favor a normal mass ordering. Our results show that the constraints on the parameter space that come from LFV can be stronger than the ones from colliders depending on the vacuum expectation value of the neutral field in the scalar triplet, $v_{s11}$. In particular, we derived $M_{S^{\pm\pm}}> 3$~TeV using the SINDRUM report on $\mu \rightarrow 3e$ decay for $\kappa=0$ and $\kappa=1$ (for a dominated type II scenario and a mix type I+II scenario). For a dominant type I seesaw, LFV bounds do not constraint significantly the parameter space of the model. Our conclusions hold regardless of the neutrino oscillation parametrizations adopted and the precise value of the sum of the neutrino masses.

\section{Acknowledgments}
We thank Gabriela Hoff and Gabriela Lichtenstein for discussions.  ANID - Millennium Science Initiative Program - ICN2019\_044. This work is supported by Simons Foundation (Award Number:1023171-RC), FAPESP Grant 2018/25225-9, 2021/01089-1, 2023/01197-4, ICTP-SAIFR FAPESP Grants 2021/14335-0, 
CNPQ Grants 403521/2024-6, 408295/2021-0, 403521/2024-6, 406919/2025-9, 351851/2025-9, ANID-Millennium Science Initiative Program ICN2019\_044, and IIF-FINEP grant 213/2024. 
The work of J.C.H. is supported by ANID FONDECYT regular No.1241685.
The work of T.O. is supported by ANID FONDECYT regular No.1250343.
The work of A.R. is supported by ANID FONDECYT regular No.1230110.
We thank IIP for the local cluster {\it bulletcluster} which was instrumental in this work.

\bibliographystyle{JHEPfixed}
\bibliography{literature331}
\end{document}